# Finite Element Analysis of Biomechanical Interactions of a Subcutaneous Suspension Suture and Human Face Soft-Tissue: A Cadaver Study

Seyed Ali Mousavi[1,2,3], Mohammad Ali Nazari[1,2*], Pascal Perrier[3], Masoud Shariat Panahi[1], John Meadows[4], Marie-Odile Christen[4], Ali Mojallal[5], Yohan Payan[2]

[1]*Department of Mechanical Engineering, College of Engineering, University of Tehran, Tehran, Iran*

[2]*Univ. Grenoble Alpes, CNRS, Grenoble-INP, TIMC, Grenoble, France*

[3]*Univ. Grenoble Alpes, CNRS, Grenoble-INP, GIPSA-LAB, Grenoble, France*

[4]*Sinclair Pharmaceutical Ltd, Chester, UK*

[5] *Department of Plastic and Adhesive Surgery, Croix-Rousse Hospital, Hospices Civils de Lyon, Université Claude Bernard Lyon 1, Lyon, France*

*Corresponding Author

Mohammad Ali Nazari
```
School of Mechanical Engineering, Biomechanics Department
College of Engineering, University of Tehran
Tel: +98 (0) 21 8800 5677 -Fax: +98 (0) 21 8801 3029
Office: +98 (0) 21 6111 5237
```
*Email: manazari@ut.ac.ir*

# Finite Element Analysis of Biomechanical Interactions of a Subcutaneous Suspension Suture and Human Face Soft-Tissue

**Abstract**: In order to study the local interactions between facial soft-tissues and a Silhouette Soft® suspension suture, a CE marked medical device designed for the repositioning of soft tissues in the face and the neck, Finite element simulations were run, in which a model of the suture was embedded in a three-layer Finite Element structure that accounts for the local mechanical organization of human facial soft tissues. A 2D axisymmetric model of the local interactions was designed in ANSYS, in which the geometry of the tissue, the boundary conditions and the applied loadings were considered to locally mimic those of human face soft tissue constrained by the suture in facial tissue repositioning. The Silhouette Soft suture is composed of a knotted thread and sliding cones that are anchored in the tissue. Hence, simulating these interactions requires special attention for an accurate modelling of contact mechanics. As tissue is modelled as a hyper-elastic material, the displacement of the facial soft tissue changes in a nonlinear way with the intensity of stress induced by the suture and the number of the cones. Our simulations show that for a 4-cone suture a displacement of 4.35 mm for a 2.0 N external loading and of 7.6 mm for 4.0 N. Increasing the number of cones led to the decrease in the equivalent local strain (around 20%) and stress (around 60%) applied to the tissue. The simulated displacements are in general agreement with experimental observations.

# 1. Introduction

For many years, aesthetic clinical procedures addressing facial ptosis, which is characterized by the drooping of facial soft tissues, have typically involved invasive face lift surgery. These surgeries often require sedation under general anaesthesia and may result in tissue lesions, scars, pain, and prolonged recovery periods. (see [29] or [16]). In order to reduce the severity of the procedure and avoid these potential sequels, Sulamanidze et al. (2002) [30] have proposed a minimally invasive tissue repositioning technique that involves the insertion of suspension threads, or *sutures*, in the superficial face tissues.

In recent years, these minimally invasive aesthetic procedures have become increasingly popular for patients not wanting to undergo face lift surgeries. Suspension sutures are medical devices in which a series of structural elements such as barbs or cones are attached to the thread. These elements are designed to physically engage with the surrounding tissues thereby anchoring the device in place, after their implantation via needle insertion techniques into the subcutaneous fat pads. This allows the clinician to apply tension to the implanted device and to move parts of the facial tissues to a new position and fix them efficiently in their new position by anchoring a more distant part of the device into a second facial region made of less mobile tissue. Because they do not imply aggressive suspension, sutures can be employed in for correction of facial paralysis [1,5,8] and for aesthetic purposes to enhance the appearance of the midface [17, 20, 21,28]. Flynn et al. [12] provided a comprehensive review of the different methods involving suture suspensions.

Silhouette Soft [24] is a CE marked suspension suture, made of bioresorbable polymers, which incorporates a series of hollow cones (Figure 1). The suture is introduced, thanks to surgical needles, inside the hypodermis layer, along a predetermined pathway which is drawn on the face before the insertion after the clinician's assessment of the desired tissue repositioning to be achieved. The role of the hollow cones is to ensure the anchoring of the suture in the tissues. The clinical use and material properties of these sutures have been studied in the literature for example see [6,9,10,11,21].

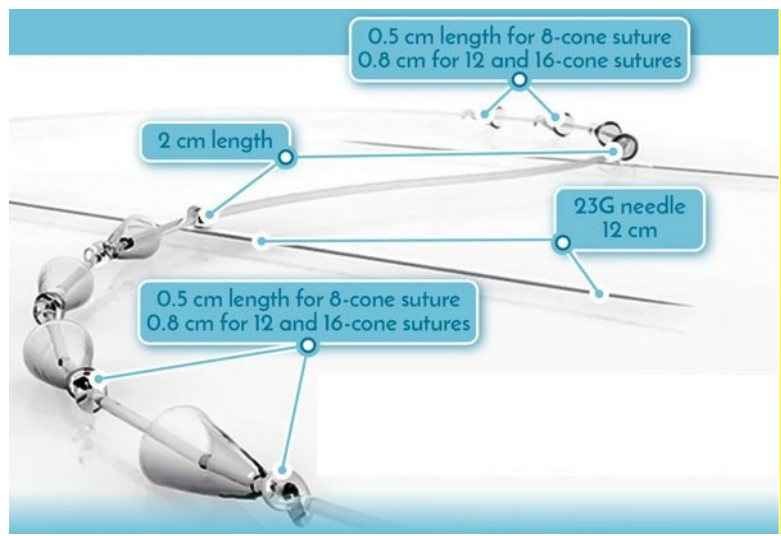

**Fig. 1.** Silhouette Soft sutures geometry and schematic view (from https://www.sinclairpharma.com/

Having a special geometrical design, the mechanical consequences of various characteristics of the suture, e.g. the type of conic part (hollow or full), or the number of conic parts (1 to 4), are important issues to be investigated. Moreover, it is desirable to assess the results obtained with this suture based on quantitative data rather than just qualitative visual comparisons. So, it is important to study the biomechanical interactions of the suture-tissue coupling numerically.

Finite Element (FE) modelling has been shown to be an efficient way to investigate biomechanical problems, in the face [3,4,13,14,15] or the tongue [27]. Interactions between barbed suture and soft tissue were studied by Ingle et al. [18, 19], who considered a linear soft tissue model. This linear assumption is consistent with their experiment which was tested on nearly stiff bovine animal skin and in a small range of strains, but in this work we consider nonlinearity for human tissue it is significantly softer.

This paper presents the results of an evaluation of the impact of the number of cones in the sutures on face lift. In this aim a model of the suture has been developed and inserted in a Finite Element cylindrical model that approximates the human face soft tissues. This cylinder has been modelled as a three-layer medium mimicking the complex structure of the skin that includes the Epidermis, Dermis, and Hypodermis layers with a nearly incompressible hyperelastic constitutive behaviour that takes into account the nonlinearity of these tissues. An assessment of the results based on a 'pull-out' study of the suture in a fresh human cadaver has been performed.

## 2. Results

The physical interaction between the suture and the face soft-tissues is mainly related to the anchorage of the cones in the tissue and the capacity of the cones to exert a stress on the tissues. Using the FE method, we modelled the suture embedded in a sample of soft tissue, taking into consideration the non-linear mechanical characteristics of the soft tissue. Different numbers of conic parts, i.e. 1, 2, 3 or 4 cones, and different loadings including 1, 2, 3 and 4 N were considered in the simulations. The evaluation of the simulations is based on 3 quantities: the maximum of the sum of the displacement components (USUM), corresponding to the maximum total displacement, the maximum of the equivalent stresses (SEQV), which gives the maximum von-Mises stress, and the maximum of total equivalent elastic strain (EPTOEQV), which gives the maximum von-Mises strain.

## 2.1. Tissue-suture interactions with a 4-cone suture

The results obtained with the 4-cone model for 1.0, 2.0, 3.0 and 4.0 N loadings are reported in Table 1.

**Table 1**. Maximum total displacement and equivalent stress and strain results for the 4-cone model

| Applied Force (N) | 1 | 2 | 3 | 4 |
|---|---|---|---|---|
| USUM (mm) | 4.0 | 5.3 | 6.7 | 7.6 |
| SEQV.(kPa) | 23549 | 587.78 | 1100 | 2450 |
| EPTOEQV. | 1.24 | 1.57 | 1.85 | 2.03 |

For a more precise evaluation, Fig. 2 shows the USUM distributions in the 1.0, 3.0 and 4.0 N cases.

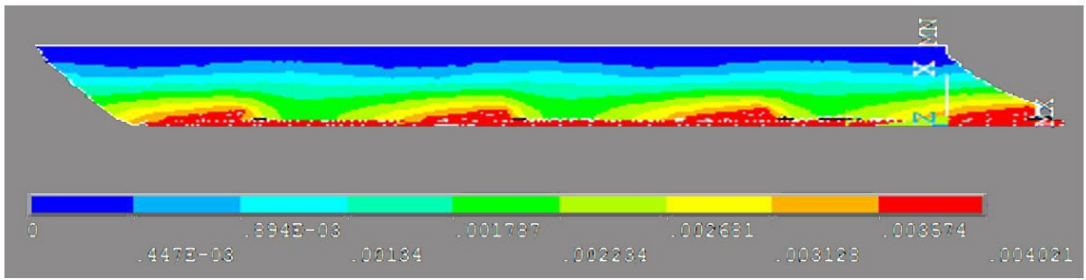

(a)

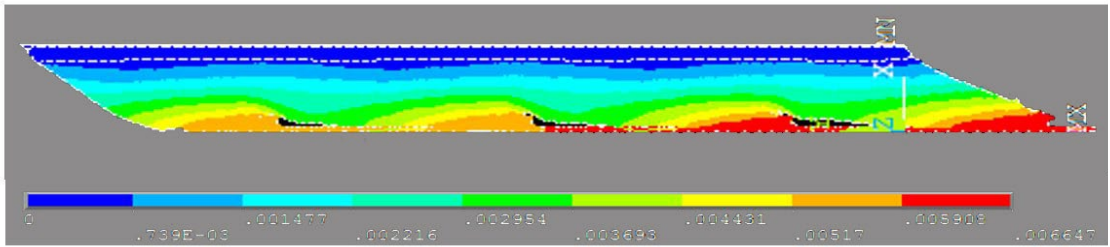

(b)

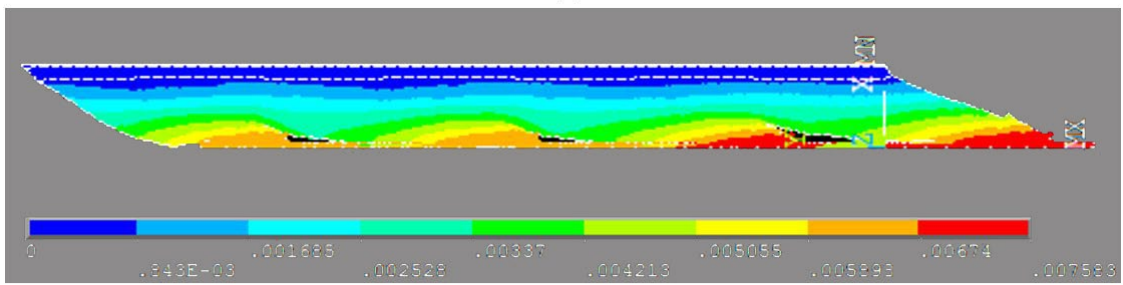

(c)

**Fig. 2.** (a) 1.0 N 4-cone case- USUM contours; (b) 3.0 N 4-cone case- USUM contours; (c) 4.0 N 4-cone case- USUM contours (USUM in m).

## 2.2. Effect of the number of cones in the suture

In order to study the effects of the number of cones on the USUM, simulations were run under the same conditions as above, with 1, 2 and 3- cone models. Fig. 3 represents the corresponding USUM distributions for a 1N pulling force.

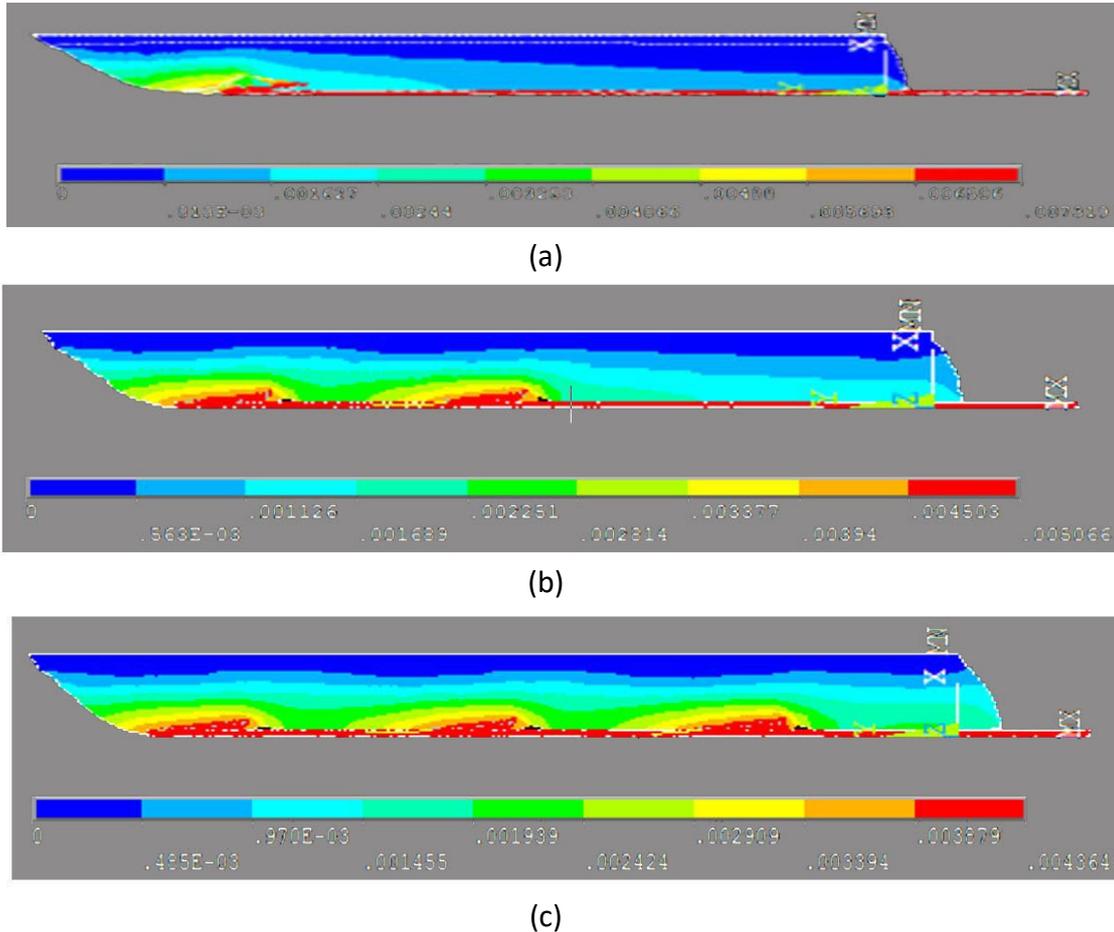

(a)

(b)

(c)

**Fig. 3.** (a) 1 cone model applying 1.0 N force- USUM contours; (b) 2-cone model applying 1.0 N force- USUM contours; (c) 3-cone model applying 1.0 N force- USUM contours- (USUM in m).

The comparison of the three panels of Fig. 3 with Fig. 2a shows two important results: (1) the USUM reached in the tissue model is the strongest for 1 cone (7.3 mm) and it continuously decreases when the number of cones increases (2 cones: 5.1 mm; 3 cones: 4.4 mm; 4 cones: 4.0 mm); this is associated with strong differences in SEQV and EPTOEQV (Table 2); (2) the overall tissue displacement is stronger and more smoothly distributed within the tissue, when the number of cones increases, as evidenced by the large green area, corresponding to a 2 mm USUM, observed for 3 and, even better, for 4 cones, and by the fact that in the 4-cone configuration the USUM is roughly the same for each cone and for each portion of tissue surrounding a cone. Finally, Fig. 2c shows that a pulling force of 4.0 N is required to generate in the 4-cone thread a USUM of 7.6 mm, which is close the one measured for 1.0 N with the 1-cone thread. The counterpart to this greater force is a better distribution of displacement in the whole piece of tissue.

**Table 2.** Maximum of equivalent stress and strain results for different numbers of conic parts with a 1N pulling force

| Model | 2-cone | 3-cone | 4-cone |
|---|---|---|---|
| SEQV.(Pa) | 662340 | 448324 | 235490 |
| EPTOEQV. | 1.57 | 1.42 | 1.24 |

### *2.3. Validation on a cadaver*

Measurements on cadaver are compared with model data (Fig. 4). The results show a maximum 18 percent error. The error stems from assuming the mechanical properties of the model based on reported values from the literature, which are not the actual mechanical properties of the current cadaver. This highlights the importance of accurately estimating the mechanical properties of facial tissue in a patient specific manner, as it is a requirement for a close estimation of suture behaviour on the subject.

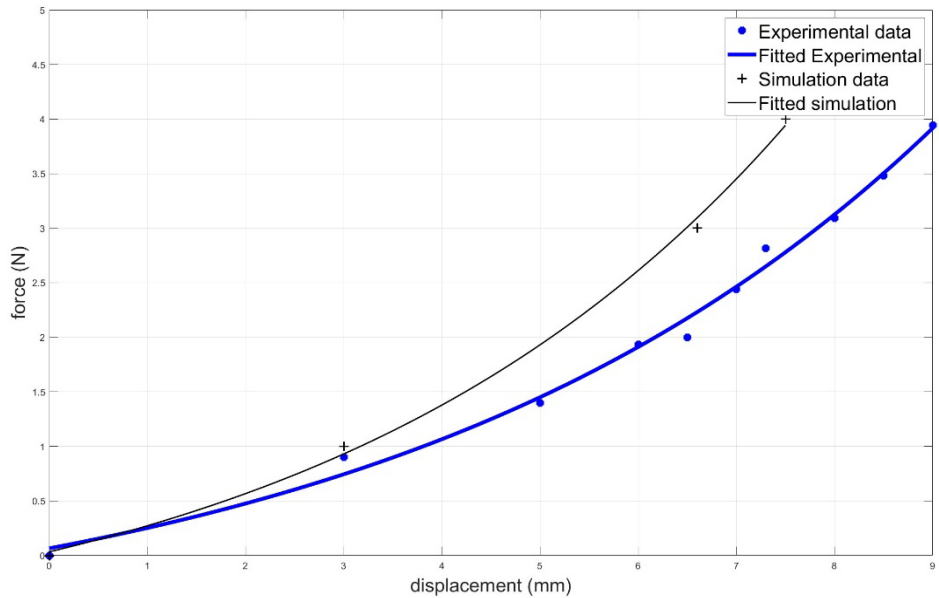

Fig. 4 Force vs suture end displacement for experimental measurement (blue curve) and simulation data (black curve)

## 3. Discussion

Comparisons were made between models having different numbers of cones. They show that, for a given force applied to the thread the global displacement of the soft tissue is more evenly distributed throughout the tissue when the thread has 4 cones (Fig. 2a and Fig. 3). Thus 4-cone threads avoid the creation of localized compression points around the cone, where a strong stress is exerted inducing a strong deformation (see Table 2), two physical phenomena that are likely to deteriorate the imposed tissues. This observation speaks in favour of threads with a large number of cones. However, this very positive feature comes at a price: for the same level of pulling force the magnitude of the largest displacement with a 4-cone thread is around 60% of the one obtained with a 1-cone thread. The increase of the number of contact regions with the number of cones induces a larger resistance of the tissues to displacement, as the applied energy distributes in more interactional regions. Hence, when they insert the thread, clinicians should proceed in order to induce in the thread a level of tension that is adapted to the desirable extent of tissue repositioning. Also, our results exhibit a nonlinear trend, both in the variations of displacements, stresses, and strains with the amount of applied force (Table 1) and with the number of cones (Table 2). It can be interpreted as the nonlinear behaviour of the materials of the tissue interacting in a contact simulation.

## 4. Materials and methods

Interactions of the Silhouette Soft suture inside a block of three-layer soft tissue was investigated. FE simulations were conducted to study the effects of the number of conic parts and of the loading conditions.

### *4.1. The Silhouette Soft® suture*

Silhouette Soft® [24] (Sinclair Pharma, UK) (Fig. 1) consists of a resorbable monofilament made of Poly-L-Lactic acid (PLLA) [10], on to which are threaded a series of hollow cones made of a resorbable copolymer of Poly (L-Lactic-co-glycolic) acid in which the monomer ratio is 82%/18% lactic acid/glycolic acid respectively. The movement of each cone along the monofilament is limited between a series of evenly spaced knots tied in the monofilament, with only one cone being located between each pair of adjacent knots. The knots serve as a tool for positioning the cones. As the surgeon pulls the suture thread, the knots become tight and act as a stopping point for the cones. This action was not modelled, and the modelling studies the effect of pulling on facial tissues after the cones have been anchored in the tissue and on the thread.

. The cones are arranged in a symmetrical bidirectional configuration with the two sets of oppositely facing cones (n ≥ 4 cones per set) being separated by a central cone free section. Each cone is 2.54 mm high, with a 1.27 mm external diameter at the basis, a 0.53 mm external diameter at the top and a 0.13 mm thickness of the material.

The complex structure of the skin can be considered as a three-layer tissue including Epidermis, Dermis, and Hypodermis layers. Using surgical needles, the suture is inserted into the hypodermis layer following a predetermined pathway that is marked on the face. This pathway is determined by the clinician's assessment of the desired tissue repositioning that needs to be accomplished. To ensure that the cones fully engage with the surrounding tissues, after implantation, the clinician applies a localised and directional massage of the face to facilitate the movement of soft tissue over the cone in the direction towards its basis.

### *4.2. Modelling the interactions of the suture with face soft tissues*

*Geometrical modelling: Finite element mesh*

We have modelled this interaction with an FE model of the face soft tissue. Given the small size of the cones, and the complexity of the contact modelling, the elements of the FE mesh had to be very small (see below). In order to reduce the computational complexity associated with high density FE meshes, we assumed a 2D axisymmetric modelling framework [23, 31] (Fig. 5a), enabling a fair mechanical account of the 3D problem while relying on a 2D modelling (Fig. 5b). This is a reasonable approach, since the suture is a fully axisymmetric.

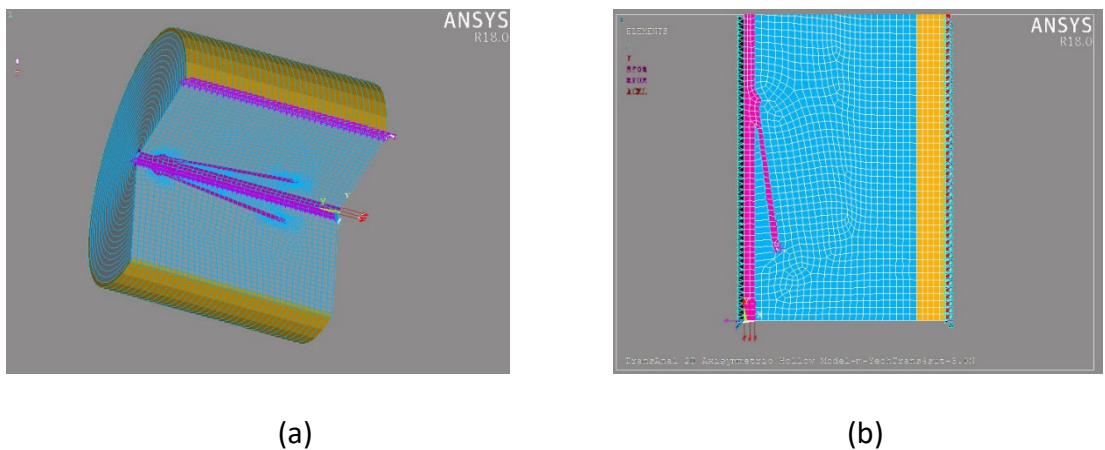

(a)          (b)

**Fig. 5.** Schematic view of 2D axisymmetric model (left panel) and its ¾ volume Expansion view (right panel).

In the *ANSYS® (Academic Research Mechanical, Release 18.2)* modelling environment, the 2D axisymmetric mesh consisted of 4-node plane 182 quadratic elements, which are basically 4-node 2D elements with axisymmetric behaviour for selected key-options. Face soft tissues were modelled in three layers. For computational issues, only part of tissue and half of the suture have been modelled (Fig. 5b). To prevent the tissue from moving freely when a force was applied to the suture, it was secured to a fixed boundary. Thus, the outer surface of the epidermis layer in the FE mesh was constrained in all degrees of freedom. Since the relative motion of suture is important, this assumption is compatible with a general assessment of the suture behaviour in the facial tissue.
.

*Mechanical modelling*

In the three-layer model of the soft tissue, the symmetrical thicknesses of hypodermis, dermis and epidermis layers were 3.0, 0.5 and 0.1 mm respectively [4]. We made investigations to find the proper constitutive laws of the tissue and suture materials. This included an experimental study and a literature survey [7, 15, 26]. The material behaviour for the suture was identified to be linear isotropic, and its properties were determined by uniaxial testing (Fig. 6). The material behaviour can be assumed to be linear isotropic in the considered range of strain, with a Young's modulus of E=1.5 GPa. The Poisson's ratio was set to 0.4, as given in [19, 31].

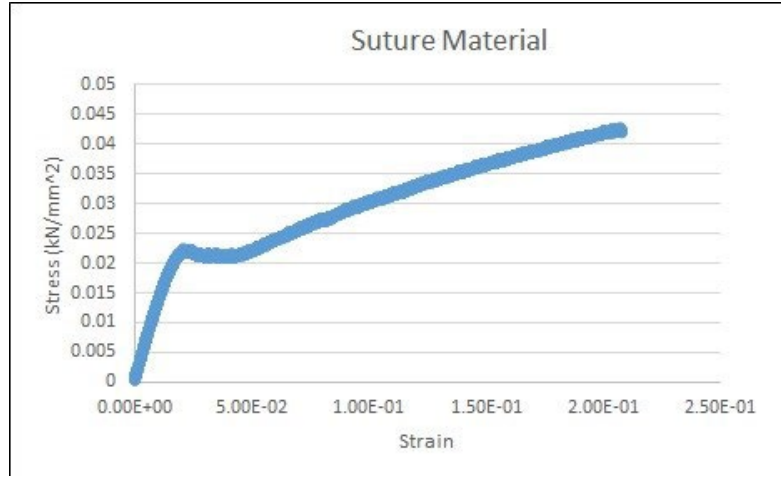

**Fig. 6:** Stress- Strain behaviour of suture material.

A two parameter Yeoh constitutive model [26] was adopted for the hypodermis and dermis, while the Gent model was considered for the epidermis. Yeoh material model accounts for a suitable nonlinear behaviour in contact simulations, and Gent model can account for the higher stiffness of the epidermis. The constitutive strain energy functions are given in Eq. 1 for Yeoh material and in Eq. 2 for Gent material [7,26]:

$$W = \sum_{i=1}^{n} C_{i0}(\bar{I}_1 - 3)^i + \sum_{k=1}^{n} \frac{1}{d_k}(J - 1)^{2k} \qquad (1)$$

$$W = -\frac{\mu J_m}{2}\ln\left(1 - \frac{\bar{I}_1 - 3}{J_m}\right) + \frac{1}{d}\left(\frac{J^2 - 1}{2} - \ln J\right) \qquad (2)$$

where $W$ is the strain energy density. In Eq.1 $C_{i0}$, are material constants, $d_k$ incompressibility parameters, $J$ the Jacobian of the elastic deformation gradient F, $\bar{I}_1$ the deviatoric first principal invariant, $\bar{I}_1 = J^{-2/3}I_1$, and $I_1$ the first invariant of right Cauchy-Green strain tensor **C=F$^T$F**. In Eq. 2 $\mu,,$, is the shear modulus, $J_m$ the asymptotic parameter, which defines the limiting value of $\bar{I}_1 - 3$, and $d$ the incompressibility parameter [2]. The coefficients used for these models are given in Table 3, taken from literature [3, 15].

**Table 3.** Coefficients of the Hyperelastic Models

| Skin Layer | Model | $C_{10}$ (Pa) | $C_{20}$ (Pa) | $1/d_1$ (kPa) | $1/d_2$ (kPa) |
|---|---|---|---|---|---|

| Hypodermis | Yeoh | 400 | 1400 | 50 | 50 |
| Dermis | Yeoh | 4000 | 14000 | 50 | 50 |
| | | $\mu\ (Pa)$ | $j_m$ | $1/d\ (kPa)$ | |
| Epidermis | Gent | 4000 | 1.2 | 50 | |

External loadings in the order of different N forces are applied at the end of the suture. Two-load step transient analysis was considered. The loading function is shown in Fig. 7. The maximum force varied from 1 to 4 N and it was held for 0.2 sec in order to stabilize the loading effects.

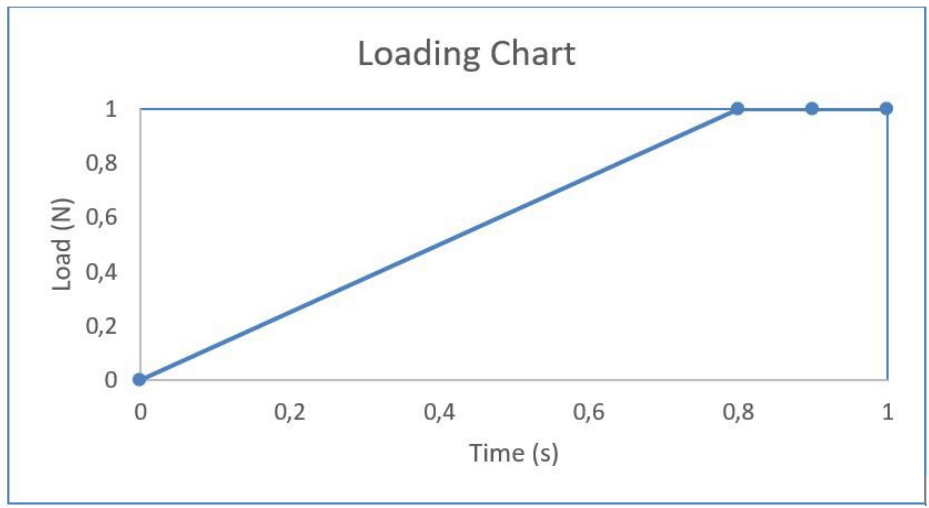

**Fig. 7.** Loading function (maximum applied forces were 1, 2, 3 or 4 N within the same time frame).

Facing convergence issues while using various contact algorithms, we selected the Multi Point Constraint (MPC) method, considering that it can be trusted since there isn't much sliding between suture and tissue after the insertion of the suture (see Appendix). The MPC contact algorithm of ANSYS [2] defines surface-to-surface contact type in the form of pair-based contact, using Targe169 target elements and Conta171 contact elements, which specifications were set using their different key options. The validation of this choice was confirmed by the strong correlation between our results and measurements taken on a cadaver. (see Results).

**4.3 Experimental measurements on fresh cadaver**

An experimental investigation of the suture-soft tissue interactions was performed on a fresh cadaver head with high skin quality (with a head size of 24x14 cm) in the anatomy lab of Rockefeller medical school (University Claude Bernard, Lyon 1). Before suture insertion the cadaver was examined for quality and uniformity of the material. A 4-cone portion of the suture was implanted. The suture was inserted in the midface (hypodermis facial layer) following recommended aesthetic procedures. The first cone was located close to the exit point corresponding to the reference mark on the skin. The external part of the suture was cut so that only a sufficient length of the proximal end remains which can be attached to the hook end of the handheld tensiometer. The tensiometer was a digital force gauge (Dynamometer PCE-DFG 20, PCE Deutschland GmbH). Various tensile forces were applied at the end of the suture and the induced relative displacements of the tissue were measured (Fig. 8).

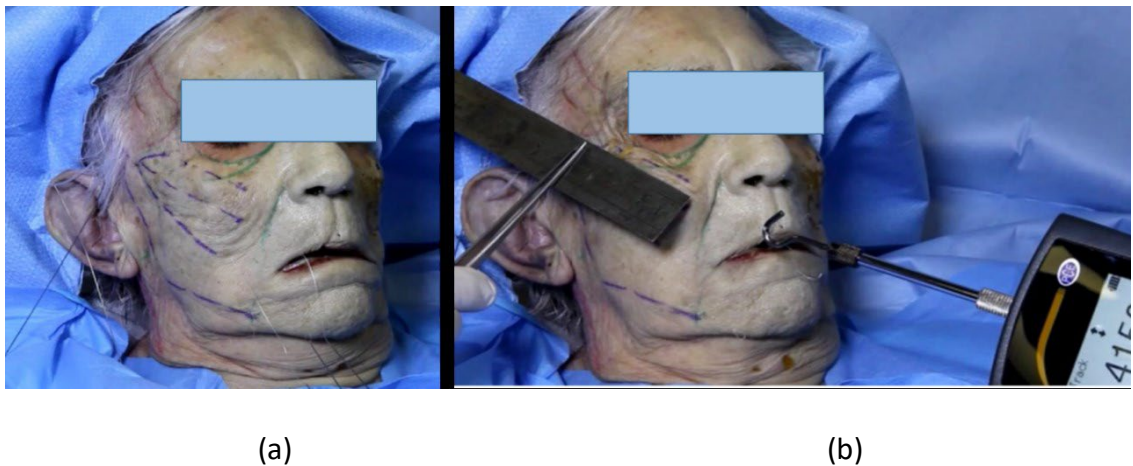

(a)                  (b)

**Fig. 8.** (a) Cadaver Test Experiment; Panel (b) Measuring the displacement in the face when applying nearly 4.0N pulling-out force.

## 5. Declaration

### *5.1. Ethics approval and consent to participate*

All experiments on cadaver has been done by AM (maxillofacial surgeon) in anatomy laboratory of l'Hôpital de la Croix Rousse de Lyon, following the standard procedures for cadaveric research.

### *5.2. Consent for publication*

Each author contributed to drafting or revising the article for significant intellectual content, provided final approval of the version, and consented to its publication.

### *5.3. Availability of data and material*

Since the experimental data were collected on human cadavers, they will not be publicly available. They will be obtained upon request and after agreement of the ethical authority of Lyon University.

### *5.4. Competing interests*

Authors have no conflicts of interest or financial ties to disclose.

### *5.5. Funding*

This work was partly funded by the Sinclair Pharmaceuticals LTD company.

### 5.6. Authors' contributions

MAN, PP, MSP, JM, MOC and YP contributed in design of the concept and project supervision. SAM contributed in performing simulations and data analysis. MOC and AM provided required data in cadaver study.

**Appendix**

In FE, after assembling element matrices and before applying the boundary conditions (BCs), the global form of governing equations becomes:

$$KX = F \qquad (1p)$$

where $K$ is the global stiffness matrix, $X$ the vector of degrees of freedom (DOF) (without any BCs), and $F$ the global force vector (including nodal forces and element forces). A general Multi Point Constraint (MPC) between a *contact* surface and a *target* surface can be written as:

$$X_t = AX_c + B \qquad (2p)$$

where $X_c$ is the vector of degrees of freedom of nodes on the contact surface and $X_t$ the vector of degrees of freedom of nodes for the corresponding target surface. $A$ and $B$ are matrix of coefficients and vector of constants respectively. As an example of bonded surfaces we set: $A=I$ ($I$ shows identity matrix) and $B$ shows the constant separation distance between corresponding points. To implement the eq. (2p) in (1p), the global vector of DOFs, $X$, are partitioned as:

$$X = [X_{nc} \ X_c \ X_t]^T \qquad (3p)$$

where $X_{nc}$ is the DOF vector of non-contact nodes. The eq. (1p) is then partitioned as:

$$\begin{bmatrix} K_{nc \times nc} & K_{nc \times c} & K_{nc \times t} \\ K_{c \times nc} & K_{c \times c} & K_{c \times t} \\ K_{t \times nc} & K_{t \times c} & K_{t \times t} \end{bmatrix} \begin{Bmatrix} X_{nc} \\ X_c \\ X_t \end{Bmatrix} = \begin{Bmatrix} F_{nc} \\ F_c \\ F_t \end{Bmatrix} \qquad (4p)$$

Replacing target DOFs from eq. (2p) into eq. (4p), therefore two equations (2p) and (4p) can be expressed in a single equation:

$$\begin{bmatrix} K_{nc\times nc} & K_{nc\times c} + K_{nc\times t}A & 0 \\ K_{c\times nc} & K_{c\times c} + K_{c\times t}A & 0 \\ 0 & -A & I \end{bmatrix} \begin{Bmatrix} X_{nc} \\ X_c \\ X_t \end{Bmatrix} = \begin{Bmatrix} F_{nc} - K_{nc\times t}B \\ F_c - K_{c\times t}B \\ B \end{Bmatrix} \quad (5p)$$

This set of equations are non symmetric, to restore symmetry the last line of equations in eq. (4p) are multiplied by $A^T$ and after adding to the second set of equations in (5p) and premultiplying the constraint equation in $-K_{t\times t}$ results to the following symmetrical equations:

$$\begin{bmatrix} K_{nc\times nc} & K_{nc\times c} + K_{nc\times t}A & 0 \\ K_{c\times nc} + A^T K_{t\times nc} & K_{c\times c} + K_{c\times t}A + A^T K_{t\times c} & A^T K_{t\times t} \\ 0 & K_{t\times t}A & -K_{t\times t} \end{bmatrix} \begin{Bmatrix} X_{nc} \\ X_c \\ X_t \end{Bmatrix} = \begin{Bmatrix} F_{nc} - K_{nc\times t}B \\ F_c - K_{c\times t}B + A^T F_t \\ -K_{t\times t}B \end{Bmatrix}$$

(6p)

The resolution of (6p) directly provides the solution [32]. MPC algorithm in ANSYS ties contact and target surfaces internally based on the contact kinematics. In this method since no penalty is associated with application of constraint (2p), therefore the contact-based results (like contact pressure) are not computed [2].